\documentclass[a4paper,fleqn,usenatbib]{mnras}
\usepackage[T1]{fontenc}
\usepackage{ae,aecompl}
\usepackage[dvipsnames]{color}
\newcommand{\revision}[1]{{           {#1}}} 
\newcommand{\half}{\frac{1}{2}}
\newcommand{\thrd}{\frac{1}{3}}

\newcommand{\dar}{\partial_r}   
\title{Theoretical implications of the galactic radial acceleration 
relation of McGaugh, Lelli, and Schombert}
\author[R. K. Nesbet]
{Robert K. Nesbet\thanks{E-mail: rkn@earthlink.net} \\
IBM Almaden Research Center, 
650 Harry Road,
San Jose, CA 95120-6099, USA}
\date{Accepted XXX. Received YYY; in original form ZZZ}
\pubyear{2018}
\begin{document}
\label{firstpage}
\pagerange{\pageref{firstpage}--\pageref{lastpage}}
\maketitle
\begin{abstract}
Velocities in stable circular orbits about galaxies, a measure of
centripetal gravitation, exceed the expected Kepler/Newton velocity
as orbital radius increases.  Standard LCDM attributes this anomaly
to galactic dark matter.  McGaugh et al have recently shown
for 153 disc galaxies that observed radial acceleration is an 
apparently universal function of classical acceleration computed for
observed galactic baryonic mass density.  This is consistent with the 
empirical MOND model, not requiring dark matter.  It is shown here that 
suitably constrained LCDM and conformal gravity (CG) also produce such  
a universal correlation function.  LCDM requires a very specific dark 
matter distribution, while the implied CG nonclassical acceleration 
must be independent of galactic mass.  All three constrained radial 
acceleration functions agree with the empirical baryonic $v^4$ 
Tully-Fisher relation.  Accurate rotation data in the nominally flat
velocity range could distinguish between MOND, LCDM, and conformal 
gravity.
\end{abstract}
\section{Introduction}
\par Velocities of objects in stable circular orbits about galaxies
measure radial gravitational acceleration.  Kepler/Newton velocity 
falls below observed velocity as orbital radius increases.
Standard LCDM attributes observed excess velocity to centripetal  
acceleration due to galactic dark matter.
\par \citet{MLS16} have recently shown for 153 disc galaxies that
observed acceleration $a$ is effectively a universal function of 
Newtonian acceleration $a_N$, computed for the observed baryonic  
distribution.
\revision{
The empirical function has negligible observed scatter.
}
\par This radial acceleration relation (RAR) is compatible
with the empirical MOND model\citep{MIL83,SAN10,FAM12,MIL16}, which 
does not invoke cold dark matter (CDM).  It implies some very simple
natural law.  It is shown here that such a law is predicted by 
conformal theory\citep{WEY18,MAN06,NES13,NESM5}, without dark matter.
It does not exclude a specific CDM source density derived here.
This is consistent with conclusions that current LCDM simulations 
do not imply observed galactic rotation curves\citep{WAK15}
nor the observed distribution of extragalactic matter\citep{KRO16}.
\revision{
Incremental nonclassical CG radial acceleration $\Delta a$, constant
except for a smooth cutoff
in the very large spherical dark halo, is determined in the isotropic
FLRW metric\citep{NESM3}.  Derivations of $\Delta a$, important only
for large orbital radii, are simplified here by imposing spherical
symmetry, valid for acceleration at large radii.
}
\par The empirical RAR\citep{MLS16} resolves a longstanding
conflict between MOND\citep{MIL83,SAN10,FAM12} and conformal
gravity.  Fitting conformal gravity to earlier less precise 
data\citep{MAN06,MAO11}, nonclassical acceleration parameter $\gamma$ 
has been inferred to depend on galactic mass, incompatible with 
\revision{
negligible scatter of the observed RAR.
}
Mass-independent $\gamma$ would agree with the RAR 
and with constant MOND scale parameter $a_0$.
This supports a recent conclusion that the assumed mass-dependent part
$\gamma_G=N^*\gamma^*$ of $\gamma$ cannot be derived from current 
theory\citep{NESM5}.
\par Agreement of conformal gravity with the RAR supports a universal 
conformal symmetry postulate\citep{NES13}, that all elementary physical 
fields satisfy local Weyl scaling symmetry\citep{WEY18},
modifying Einstein-Hilbert general relativity
(conformal gravity)\citep{MAK89,MAN06}
and the electroweak scalar field model
(conformal Higgs model)\citep{NESM1,NESM2}.

\section{Qualitative implications}
\par The observed correlation\citep{MLS16} between classical and
nonclassical centripetal acceleration puts a strong constraint on any
theoretical model.  Observed radial acceleration $a$ must be a unique 
function of Newtonian $a_N$, regardless of galactic structure or mass.
Standard LCDM assumes that a preexisting dark matter aggregate attracts 
baryonic matter, which forms the observable galaxy.  
\revision{
This must correlate the baryonic distribution to the assumed dark matter
with no interaction other than gravity.
}
\par $a$ and $a_N$ are functions of two variables, galactic mass $M$ 
and radius $r$, even beyond the range of dependence on galactic
structure.  If unrelated functions $a(x,y)$ and $b(x,y)$ were plotted
against each other the general result would be a 2-dimensional smear,
not the 1-dimensional line plot found by \citet{MLS16}.
This result requires $a(x,y)=F(b(x,y))$. Observed correlation function 
$F$ depends on only a single variable, with negligible scatter.
\par Correlation function $a=F(a_N)$ is a basic postulate of
MOND\citep{MIL83,MIL16}.  It will be shown here that conformal
gravity\citep{MAN06} and the depleted halo model\citep{NESM3} produce
such a function if the implied nonclassical acceleration parameter
is independent of galactic mass.  A particular distribution of dark
matter is derived here for which LCDM also produces such a function. 

\section{MOND background}
\par MOND\citep{MIL83,SAN10,FAM12} modifies the Newtonian force law
for acceleration below an empirical scale $a_0$. Using $y=a_N/a_0$
as independent variable\citep{MIL16,MCG08},
for assumed universal constant $a_0$,
MOND postulates an interpolation function $\nu(y)$ such that observed
radial acceleration $a=F(a_N)=a_N\nu(y)$,
which defines a correlation function.
\par For $a_N\gg a_0$, $\nu\to1$ and for $a_N\ll a_0$,
$\nu^2\to 1/y$.  This implies asymptotic limit $a^2\to a_0a_N$
for small $a_N$, which translates into an
asymptotically flat rotational velocity function $v(r)$ for
large galactic radius $r$\citep{MIL83}.

\section{LCDM background}
\revision{
\par External Schwarzschild potential function $B(r)$
is determined for a static spherical galactic model
by simplified second-order differential equation
\begin{eqnarray} \label{Beq}
 \dar^2(rB(r))=rw(r),
\end{eqnarray}
for $w(r)$ determined by source energy-momentum.
}
Centripetal radial acceleration for a stable circular orbit is
\begin{eqnarray} \label{areq}
 a(r)=\frac{v^2(r)}{r}=\half c^2B^\prime(r). 
\end{eqnarray}
\par Spherically averaged mass/energy density $w(r)$ 
is modeled by baryonic $w_0(r)$ within galactic radius $r_G$,
embedded in dark matter $w_1(r)$ within halo radius $r_H\gg r_G$. 
Then $w(r)=w_0(r)+w_1(r)$ within $r_G$. Functions $y_0=rB(r)$  
and derivative $y_1$ satisfy differential equations
\begin{eqnarray}
 \dar y_0=y_1, \dar y_1=rw(r).
\end{eqnarray}
\par Gravitational potential $B(r)$ is required to be differentiable
and free of singularities. $B(r)=\alpha-2\beta/r$ is the source-free
solution. $y_0(0)=0$ prevents a singularity at the origin.
$y_1(0)$ can be chosen to match boundary condition $\alpha=1$ at $r_H$. 
\par A solution of Eq.(\ref{Beq}) for $r\leq r_H$ is  
\begin{eqnarray} \label{rBeq1}
 y_0(r)=rB(r)=-\int_0^rwq^2dq+\alpha r-r\int_r^{r_H} wqdq
\nonumber\\
 y_1(r)=B(r)+rB^\prime(r)=\alpha-\int_r^{r_H} wqdq.
\end{eqnarray}
\par The simple form $a=a_N+\Delta a$ defines RAR correlation function 
$F(a_N)$ if $\Delta a$ is a universal constant.  Dependence on
$r$ or $M$ would produce scatter about such a function plotted as
$a=F(a_N)$\citep{MLS16}.  Eqs.(\ref{areq}) and (\ref{rBeq1}) 
imply LCDM dark matter term  
$\Delta a=\half\frac{c^2}{r^2}\int_0^r w_1q^2dq$.
Constant $\Delta a$ requires $w_1(r)=\mu/r$ where $\mu$ is a 
universal constant.
\revision{
Constant $\Delta a$ is also implied  by the quantized inertia
model\citep{MCC13,MCC17}.
}

\section{Conformal gravity background}
\par Conformal gravity (CG) modifies the metric field action integral
of standard general relativity, replacing the Einstein-Hilbert Ricci
scalar by a quadratic contraction of the conformal 
Weyl tensor\citep{MAK89,MAN06}.  Together with the conformal 
Higgs model\citep{NESM1} of dark energy, also without dark matter, 
this follows a postulate of universal conformal symmetry\citep{NES13}.
\par In spherical geometry, the static source-free
Schwarzschild potential\citep{MAK89,MAN06} is
$B(r)=-2\beta/r+\alpha+\gamma r-\kappa r^2$,
where all coefficients are constants and
$\alpha^2=1-6\beta\gamma$\citep{MAK91}.
This 4th-order CG equation adds two integration parameters 
$\gamma,\kappa$ to the 2nd-order LCDM equation.
$\gamma$ defines nonclassical radial acceleration and $\kappa$ 
determines a cutoff at the halo boundary\citep{NESM3}.
\revision{
Outside an assumed model spherical source mass,
Schwarzschild potential function $B(r)$
determines circular geodesics such that
$ra/c^2=v^2/c^2=\half rB^\prime(r)=\beta/r+\half\gamma r-\kappa r^2$.
The Kepler formula is $ra_N/c^2=\beta/r$.
}
Agreement with standard general relativity for sub-galactic
phenomena requires $\beta=GM/c^2$.  
\par Observed orbital velocities for
138 galaxies  are fitted assuming $\gamma=\gamma_0+\gamma_G$, where
$\gamma_G=N^*\gamma^*$\citep{MAN97,MAN06} for $N^*=M/M_\odot$.
Constants inferred from this rotation data are
$\gamma_0=3.06\times 10^{-28}/m$, $\gamma^*=5.42\times 10^{-39}/m$,
and $\kappa=9.54\times 10^{-50}/m^2$
\citep{MAN97,MAO11,MAO12,OAM12,OAM15,OCM17}.
\par Well inside a galactic halo boundary, $2\kappa r/\gamma$ can be
neglected.  For $\Delta a=\half\gamma c^2$ this defines RAR 
correlation function $F(a_N)=a_N+\Delta a$ if constant $\gamma$ is  
mass-independent, as indicated by a recent study\citep{NESM5}.
\par Conformal gravity fits to galactic orbital velocities
\citep{MAN06,MAO12} determine $\gamma$ directly for galactic mass $M$
after scaling by an assumed mass-to-light ratio $\Upsilon$.
$\Upsilon$ is adjusted for each galaxy to make assumed 
$\gamma=\gamma_0+N^*\gamma^*$ as consistent as possible for a set of 
galaxies, with universal constants $\gamma_0$ and $\gamma^*$.  This 
procedure has been remarkably successful for 138 
galaxies\citep{MAN97,MAO11,MAO12,OAM12}.
\par Replacing $\gamma_0$ by total $\gamma$ and eliminating $\gamma^*$
would retain the orbital rotation velocity function to good accuracy.
The practical issue is whether or not mass-to-light parameters
$\Upsilon$ could be adjusted to give mass-independent $\gamma$. The 
recent study by \citet{MLS16} strongly indicates that this is possible.  
This study, designed to reduce observational error as much as 
possible, eliminates the need to adjust $\Upsilon$ for each galaxy.  
Conformal gravity can be empirically correct only if $\gamma$ 
is mass-independent.

\section{Determination of parameter $\gamma$}
\par In the Schwarzschild metric, nonclassical
CG acceleration parameter $\gamma$ for a galaxy has been assumed
to take the form $\gamma=\gamma_0+\gamma_G$, where 
$\gamma_G=N^*\gamma^*$\citep{MAN97}, proportional to galactic mass 
$M=N^*M_\odot$ in solar mass units. 
Mass-independent $\gamma_0$ is attributed to the Hubble flow.
\par The depleted halo model\citep{NESM3} justifies this rationale
for mass-independent $\gamma_0$.  One might anticipate a second
fundamental constant $\gamma^*$, as assumed by Mannheim 
et al\citep{MAN97,MAN06,MAO12,OAM15}. The RAR\citep{MLS16} 
requires $\gamma^*=0$.  If so, $\gamma_0$ must be identified
with inferred universal constant total $\gamma$, in agreement with 
MOND constant $a_0$.
\par A galaxy of mass $M$ can be modeled by spherically averaged mass
density ${\bar\rho}_G/c^2$ within radius $r_G$, formed by condensation
of primordial uniform, isotropic matter of mass density $\rho_m/c^2$
from a sphere of large radius $r_H$\citep{NESM3}.
\par The depleted halo model\citep{NESM3} identifies the dark halo
inferred from gravitational lensing and anomalous centripetal
acceleration with this depleted sphere.
\par Given constant mean density ${\bar\rho}_G$ within $r_G$,
this model determines empty halo radius
$r_H=r_G({\bar\rho}_G/\rho_m)^\thrd$.
Empirical parameters $\gamma$ and $\kappa$ from Schwarzschild potential
$B(r)$ imply halo radius $r_H=\half\gamma/\kappa$\citep{NESM3}.
\par For the Milky Way, $r_H=33.28\times 10^{20}m=107.8kpc$, compared
with $r_G\simeq 15.0kpc$.
\par The conformal Friedmann equation\citep{NESM3}, with Friedmann 
weight parameters $\Omega_k$ and $\Omega_m$ set to zero\citep{NESM1},
fits observed Hubble function $h(t)=H(t)/H_0$,
scaled by Hubble constant $H_0$, as accurately as LCDM,
with only one free constant for redshifts 
$z\leq 1$(7.33Gyr)\citep{NESM1,NESM2}.
This determines Friedmann weights, at present time $t_0$,
$\Omega_\Lambda=0.732, \Omega_q=0.268$,
where acceleration weight $\Omega_q=\frac{{\ddot a}a}{{\dot a}^2}$
and $a(t)$ is the computed Friedmann scale factor\citep{NESM1,NESM2}.
\par A geodesic passing into the empty halo
from the surrounding cosmic background is deflected by acceleration
proportional to incremental Hubble acceleration\citep{NESM3}
$\Delta\Omega_q=
(1-\Omega_\Lambda)(0)-(1-\Omega_\Lambda-\Omega_m)(\rho_m)=
\Omega_m(\rho_m)=
\frac{2}{3}\frac{{\bar\tau}c^2\rho_m}{H^2_0}$ \citep{NESM1}.
\par Converted from Hubble units, this implies nonclassical centripetal
acceleration $\half\gamma c^2=-cH_0\Omega_m(\rho_m)$\citep{NESM3}.
$\Omega_m<0$ because nonclassical constant ${\bar\tau}<0$\citep{NESM1}.
This is observed as gravitational lensing and in anomalous orbital
rotation velocities.
\par  This logic is equivalent to requiring continuous radial
acceleration across halo radius $r_H$ as a boundary condition:
\begin{eqnarray}
 \half\gamma_H c^2-cH_0\Omega_q(0)=-cH_0\Omega_q(\rho_m).
\end{eqnarray}
Notation $\gamma_H$ is used here for the contribution to total
acceleration parameter $\gamma$ arising from the halo boundary.
Signs here follow from the definition of $\Omega_q$ as centrifugal
acceleration weight.
\par Comparison of conformal theory with observed data depends on
exact solutions of the field equations in highly symmetric geometries
characterized by two different relativistic metrics.
The conformal Higgs model\citep{NESM1,NES13} has an exact
time-dependent, spatially uniform solution in the FLRW metric, which
describes Hubble expansion.
Conformal gravity\citep{MAK89,MAN06} has an exact solution
for spherical symmetry in the static Schwarzschild metric,
which describes anomalous galactic rotation.
\par  The equations are decoupled\citep{NESM5} by separating source
mass/energy density $\rho$ into uniform average density ${\bar\rho}$
for the conformal Higgs model and residual density 
${\hat\rho}=\rho-{\bar\rho}$ for conformal gravity.
These solutions must be made consistent by choice of parameters
and boundary conditions\citep{NESM5}.
\par For a spherical solar mass isolated in a galactic halo,
$\gamma^*=0$ results from requiring continuous radial
acceleration across boundary radius $r_\odot$\citep{NESM5}.
Mean internal mass density ${\bar\rho}_\odot$ within $r_\odot$
determines an exact solution of the conformal Higgs gravitational
equation\citep{NESM1,NES13}, giving internal acceleration weight
$\Omega_q({\bar\rho}_\odot)$.
For continuous radial acceleration across $r_\odot$,  
\begin{eqnarray} \label{rGeq}
 \half\gamma_{\odot,in}c^2-cH_0\Omega_q({\bar\rho}_\odot)=
 \half\gamma c^2-cH_0\Omega_q(0),
\end{eqnarray}
constant $\gamma_{\odot,in}$ is determined by local mean source
density ${\bar\rho}_\odot$, valid inside $r_\odot$.
$\gamma$ is a constant of integration
that cannot vary within the source-free halo.
Eq.(\ref{rGeq}) does not determine a mass-dependent increment.
\par  Thus galactic $\gamma$ consists entirely of constant $\gamma_H$
determined at halo boundary $r_H$.  It is constant and spatially 
uniform in the source-free space because it depends only on uniform 
cosmic background density $\rho_m$ and on Hubble constant
$H_0=2.197\times 10^{-18}/s$\citep{PLC15}.

\section{The Tully-Fisher relation}
\par Static spherical geometry defines Schwarzschild
potential $B(r)$.  For a test particle in a stable exterior
circular orbit with velocity $v$ the centripetal acceleration is
$a=v^2(r)/r=\half B'(r)c^2$.  Given $\beta=GM/c^2$,  
Newtonian $B(r)=1-2\beta/r$ for sufficiently large $r$,
so that $a_N=\beta c^2/r^2=GM/r^2$ .
\par MOND postulate $a^2\to a_N a_0$ as $a_N\to 0$\citep{MIL83,MCG11}
implies $v^4=a^2r^2\to GMa_0$.  This supports the empirical baryonic
Tully-Fisher relation\citep{TAF77,MCG05,MCG11}
\par CG function $B(r)$ determines orbital velocity in the source-free
halo $v^2/c^2=ra/c^2=\beta/r+\half\gamma r-\kappa r^2$. For $r$ in a
range outside $r_G$ such that Newtonian $ra_N/c^2\simeq\beta/r$,
while $2\kappa r/\gamma$ can be neglected, the slope of
$v^2(r)$ vanishes at $r^2_{TF}=2\beta/\gamma$.  This implies that
$v^4(r_{TF})/c^4=(\beta/r_{TF}+\half\gamma r_{TF})^2=
  2\beta\gamma$\citep{MAN97,NESM5}.
This is the Tully-Fisher relation, exact at stationary point
$r_{TF}$ of the $v(r)$ function.  Given $\beta=GM/c^2$,
$v^4\simeq 2GM\gamma c^2$, for relatively constant $v(r)$ centered
at $r_{TF}$.  This derivation holds for CG neglecting $\kappa$ and
for equivalent LCDM with source density $\mu/r$.  MOND $v^4=GMa_0$  
would be identical if $a_0=2\gamma c^2$\citep{MAN97}.
$r_{TF}$ can be defined as the outermost crossing point
of the Newtonian and nonclassical acceleration functions.

\section{Data for Milky Way}
\par Given $kpc=0.30857\times 10^{20}m$,
$G=6.674\times 10^{-11}m^3/kg s^2$,
$c^2=8.982\times 10^{16}m^2/s^2$,
$\gamma=6.35\times 10^{-28}/m$,
and $M=1.207\times 10^{41}kg$\citep{MAN97,MAN06,MAO11,OAM15,MCG08},
then $\beta c^2=GM=8.056\times 10^{30}m^3/s^2$.
\par Milky Way Tully-Fisher radius $r_{TF}=17.2 kpc$, halo radius
$r_H=107.8kpc$\citep{NESM3,NESM5}, for $r_G\simeq 15.0 kpc$.
Implied MOND constant $a_0=2\gamma c^2=1.14\times 10^{-10} m/s^2$.
\revision{
\par Outside $r_G$, $a_N\simeq\beta c^2/r^2$.
}
a(CDM)$=a_N+\half\gamma c^2$, 
using empirical CG $\Delta a$,
a(CG)$=a_N+\half\gamma c^2(1-r/r_H)$, including parameter $\kappa$, and
a(\citep{MLS16})$\simeq
 a_N/(1-e^{-\sqrt{a_N/a_0}})$,
just MOND with a particular interpolation function 
and $a_0=1.20\times 10^{-10}m/s^2$.
\revision{
The CDM function is generic for any model with universal constant
$\Delta a$.
}

\begin{table}
\caption{Milky Way: radial acceleration($10^{-10}m/s^2$)} \label{Tab01}
\begin{tabular}{lccccccc}
r  &     &CDM&&CG&&MOND&\\
kpc&$a_N$& a     &$10^3\frac{v}{c}$
         & a     &$10^3\frac{v}{c}$
         & a     &$10^3\frac{v}{c}$\\
\hline
15& 0.376& 0.661& 0.584& 0.621& 0.566& 0.877& 0.672\\
20& 0.212& 0.497& 0.584& 0.444& 0.552& 0.617& 0.650\\
25& 0.135& 0.420& 0.601& 0.354& 0.551& 0.475& 0.638\\
30& 0.094& 0.379& 0.625& 0.300& 0.556& 0.385& 0.630\\
35& 0.069& 0.354& 0.652& 0.261& 0.560& 0.324& 0.624\\
40& 0.053& 0.338& 0.682& 0.232& 0.565& 0.279& 0.619\\
45& 0.042& 0.327& 0.717& 0.208& 0.567& 0.246& 0.616\\
50& 0.034& 0.319& 0.740& 0.187& 0.566& 0.219& 0.613
\end{tabular}
\end{table}
\par Table \ref{Tab01} compares detailed predictions for the implied
nearly flat external orbital velocity curve for the Milky Way
galaxy.  The CDM curve rises gradually, the CG curve remains 
remarkably flat, while the MOND\citep{MLS16} curve falls gradually
toward a definite asymptotic velocity.  

\section{Conclusions}
\par LCDM, restricted to CDM source density $\mu/r$; CG, restricted to
mass-independent nonclassical acceleration parameter $\gamma$; and 
MOND, with a particular implied interpolation function\citep{MLS16}, 
are consistent with the recent RAR\citep{MLS16} and with other
qualitative features of observed stellar-dominated galactic
orbital velocities.  Velocities exceed the Newtonian value but
remain nearly constant for a large range of radii extending into
the galactic dark halo.  This constant velocity is characterized by
the baryonic Tully-Fisher relation\citep{MCG05,MCG11}, with $v^4$
proportional to baryonic galactic mass $M$.  Note that the integrated 
CDM source density produces a mass-independent constant, consistent
with CG nonclassical acceleration $\gamma$. 
\par If $\gamma$ is independent of galactic mass, CG is compatible 
with the RAR\citep{MLS16}. This supports the conclusion
that CG determines only mass-independent $\gamma$\citep{NESM5}.
Dark matter source density $\mu/r$ would determine constant
$\Delta a$ in LCDM, with the same implications as CG except at large 
radii, where CG implies effects not described by LCDM or MOND.
CG orbital velocity drops to zero at an outer boundary\citep{MAO11},
identified as the dark halo radius\citep{NESM3}.  CG parameter $\kappa$,
consistent with the halo radius, does not have a counterpart in 
LCDM or MOND.  Distinction between LCDM, CG, and MOND
requires accurate rotational data at large galactic radii.  
\par The author is grateful to colleagues Barbara Jones and John
Baglin for helpful comments.

\bsp	
\label{lastpage}

\begin{thebibliography}{99}  
\bibitem[\protect\citeauthoryear{Famaey and McGaugh}{2012}]
{FAM12} Famaey B., McGaugh S.S., 2012,
{\it Living Reviews in Relativity} {\bf 15}, 10
\bibitem[\protect\citeauthoryear{Kroupa}{2012}]
{KRO16} Kroupa P., 2012,  
The observed spatial distribution of matter.
(arXiv:1610.03854[astro-ph.CO]).
\bibitem[\protect\citeauthoryear{Mannheim}{1997}]
{MAN97} Mannheim P.D., 1997, 
{\em Astrophys.J.} {\bf 479}, 659
\bibitem[\protect\citeauthoryear{Mannheim}{2006}]
{MAN06} Mannheim P.D., 2006, 
{\em Prog.Part.Nucl.Phys.} {\bf 56}, 340
\bibitem[\protect\citeauthoryear{Mannheim and Kazanas}{1989}]
{MAK89} Mannheim P.D., Kazanas D., 1989, 
{\it ApJ} {\bf 342}, 635
\bibitem[\protect\citeauthoryear{Mannheim and Kazanas}{1991}]
{MAK91} Mannheim P.D., Kazanas D., 1991,
{\it Phys.Rev.D} {\bf 44}, 417
\bibitem[\protect\citeauthoryear{Mannheim and O'Brien}{2011}]
{MAO11} Mannheim P.D., O'Brien J.G., 2011,
{\em Phys.Rev.Lett.} {\bf 106}, 121101
\bibitem[\protect\citeauthoryear{Mannheim and O'Brien}{2012}]
{MAO12} Mannheim P.D., O'Brien J.G., 2012,
{\em Phys.Rev.D} {\bf 85}, 124020
\revision{
\bibitem[\protect\citeauthoryear{McCulloch}{2013}]
{MCC13} McCulloch M.E., 2013,
{\it EPL} {\bf 101}, 59001 
\bibitem[\protect\citeauthoryear{McCulloch}{2017}]
{MCC17} McCulloch M.E., 2017,
{\it ApSS} {\bf 362}, 149 
}
\bibitem[\protect\citeauthoryear{McGaugh}{2005}]
{MCG05} McGaugh S.S., 2005,  
{\em Phys.Rev.Lett.} {\bf 95}, 171302 
\bibitem[\protect\citeauthoryear{McGaugh}{2008}]
{MCG08} McGaugh S.S., 2008,
{\it ApJ} {\bf 683}, 137 
\bibitem[\protect\citeauthoryear{McGaugh}{2011}]
{MCG11} McGaugh S.S., 2011,
{\em Phys.Rev.Lett.} {\bf 106}, 121303 
\bibitem[\protect\citeauthoryear{McGaugh et al}{2016}]
{MLS16} McGaugh S.S., Lelli F., and Schombert J.M., 2016,
{\it Phys.Rev.Lett.} {\bf 117}, 201101 
\bibitem[\protect\citeauthoryear{Milgrom}{1983}]
{MIL83} Milgrom M., 1983,
{\it ApJ} {\bf 270}, 571 
\bibitem[\protect\citeauthoryear{Milgrom}{2016}]
{MIL16} Milgrom M., 2016, 
(arXiv:1609.06642[astro-ph.GA]).
\bibitem[\protect\citeauthoryear{Nesbet}{2010}]
{NESM2} Nesbet R.K., 2010, 
Dark energy density predicted and explained.
(arXiv:1004.5097v6 [physics.gen-ph]).
\bibitem[\protect\citeauthoryear{Nesbet}{2011}]
{NESM1} Nesbet R.K., 2011,
Cosmological Implications of Conformal Field Theory.
{\em Mod.Phys.Lett.A} {\em 26}, 893 
(arXiv:0912.0395v3 [physics.gen-ph]).
\bibitem[\protect\citeauthoryear{Nesbet}{2013}]
{NES13} Nesbet R.K., 2013, 
Conformal Gravity: Dark Matter and Dark Energy.
{\em Entropy} {\bf 15}, 152 
\bibitem[\protect\citeauthoryear{Nesbet}{2014}]
{NESM5} Nesbet R.K., 2014, 
Conformal gravity in the Schwarzschild metric.
(arXiv:1410.8076v3[physics.gen-ph]).
\bibitem[\protect\citeauthoryear{Nesbet}{2015}]
{NESM3} Nesbet R.K., 2015,
Dark galactic halos without dark matter.
{\it Europhys.Lett.} {\bf 109}, 59001 
(arXiv:1109.3626v5[physics.gen-ph]).
\bibitem[\protect\citeauthoryear{O'Brien and Mannheim}{2012}]
{OAM12} O'Brien J.G., Mannheim P.D., 2012,
Fitting Dwarf Galaxy Rotation Curves with Conformal Gravity.
{\em Mon.Not.R.Astron.Soc.} {\bf 421}, 1273
\bibitem[\protect\citeauthoryear{O'Brien and Moss}{2015}]
{OAM15} O'Brien J.G., Moss R.J., 2015, 
Rotation curve for the Milky Way galaxy in CG.
{\it J.Phys.Conf.} {\bf 615}, 012002 
\bibitem[\protect\citeauthoryear{O'Brien et al}{2017}]
{OCM17} O'Brien J.G., Chiarelli T.L., and Mannheim P.D., 2017,
(arXiv:1704.03921[astro-ph.GA]).
\bibitem[\protect\citeauthoryear{Planck Collab.}{2015}]
{PLC15} Planck Collab., 2015,
(arXiv:1502.01589v2[astr-ph.CO]).
\bibitem[\protect\citeauthoryear{Sanders}{2010}]
{SAN10} Sanders R.H., 2010,
{\it The Dark Matter Problem},
(Cambridge Univ. Press, New York).
\bibitem[\protect\citeauthoryear{Tully and Fisher}{1977}]
{TAF77} Tully R.B., Fisher J.R., 1977,
A new method for determining the distances to galaxies.
{\em Astron.Astrophys.} {\bf 54}, 661 
\bibitem[\protect\citeauthoryear{Weyl}{1918}]
{WEY18} Weyl H., 1918,
{\it Sitzungber.Preuss.Akad.Wiss.} {}, 465;
{\it Math.Zeit.} {\bf 2}, 384.
\bibitem[\protect\citeauthoryear{Wu and Kroupa}{2015}]
{WAK15} Wu X., Kroupa P., 2015,
Galactic rotation curves and the dark halo mass relation.
{\em MNRAS} {\bf 446}, 330 
\end{thebibliography}
\end{document}